\documentclass[aps,prb,twocolumn,floats]{revtex4}

\usepackage{ifthen}
\usepackage{ifpdf}
\usepackage{latexsym}
\usepackage{amsmath}
\usepackage{amssymb}
\usepackage{bm}

\ifpdf
\usepackage{graphicx}
\usepackage{epstopdf}
\else
\usepackage{graphicx}
\usepackage{epsfig}
\fi

\newcommand{\signC}{^{\!\!c}}
\newcommand{\signA}{^{\!\!a}}
\newcommand{\citeref}[1]{[\onlinecite{#1}]}

\newcommand{\im}{\mbox{Im}}

\newcommand{\eexp}{\mbox{e}^}

\newcommand{\mass}{\mathsf{m}}

\newcommand{\tbox}[1]{\mbox{\tiny #1}}

\newcommand{\amatrix}[1]{\begin{matrix} #1 \end{matrix}}

\newcommand{\be}[1]{\begin{eqnarray}\ifthenelse{#1=-1}{\nonumber}{\ifthenelse{#1=0}{}{\label{e#1}}}}
\newcommand{\ee}{\end{eqnarray}} 

\newcommand{\hide}[1]{}

\begin{document} 

\title[Transport in a double well system]
{Quantum dynamics and transport in a double well system} 

\author{Itamar Sela and Doron Cohen}

\affiliation{
Department of Physics, Ben-Gurion University, Beer-Sheva 84005, Israel}

\begin{abstract}
The simplest one-dimensional model for the studying 
of non-trivial geometrical effects is a ring shaped device 
which is formed by joining two arms. 
We explore the possibility to model such a system  
as a two level system (TLS). 
Of particular interest is the analysis of {\em quantum~stirring}, 
where it is not evident that the topology 
is properly reflected within the framework of the TLS modeling.
On the technical side we provide 
a practical ``neighboring level" approximation 
for the analysis of such quantum devices, 
which remains valid even if the TLS modeling does not apply.  
\end{abstract}

\maketitle

%%%%%%%%%%%%%%%%%%%%%%%%%%%%%%%%%%%%%%%%%%%%%%%%%%%%%%%
%%%%%%%%%%%%%%%%%%%%%%%%%%%%%%%%%%%%%%%%%%%%%%%%%%%%%%%
\section{Introduction}
\label{S1}

In this paper we explore the possibility to model
a ring shaped device (Fig.1(a)) as a two level system (TLS) (Fig.1(b)).  
We shall see that both technical and conceptual
difficulties are involved.  
The model Hamiltonian is
\be{1}
{\cal H} = \frac{1}{2\mass}\hat{p}^2+
V_A(\hat{x}-x_A)+V_B(\hat{x}-x_B)
\ee
with periodic boundary conditions 
over $x\in[-L/2,L/2]$ so as to have a ring geometry,
as illustrated in Fig.1(a).
$V_A$ and $V_B$ represent high barriers, 
such that the ring is composed of two weakly coupled arms.
We assume that we have control over some geometrical parameters
of the model and in particular over the heights $X_A$ and $X_B$ 
of both barriers. 
Our main interest is in the current that 
flows in the system. The current through an arbitrary point $x_0$ is   
obtained as the expectation value of the operator
\be{2}
{\cal I} = \frac{1}{2\mass}
\left(
\hat{p} \, \delta(\hat{x}-x_0)+
\delta(\hat{x}-x_0) \, \hat{p}
\right)
\ee

Having defined the system and its observables 
we can consider various dynamical scenarios
such as coherent {\em Bloch~oscillations} between the two arms.
Then we can ask whether a TLS modeling
is meaningful.  Of particular interest for us is
the analysis of {\em quantum~stirring} \citeref{pmx}: 
this means to induce a circulating current by periodic modulation
of the potential.

We note that transport due to periodic modulations 
of the potential~\citeref{Thouless} has been 
studied mainly in the context of quantum pumping~\citeref{bpt,BPT2,Avron}, 
where the current is induced between reservoirs.  
The notion of quantum stirring relates 
to closed geometry, where the emerging physical picture
is significantly different~\citeref{pMB,pms}.

The quantum stirring problem highlights
an obvious topological subtlety: one wonders whether
the non trivial topology of the ring is properly 
reflected in the effective TLS model.

On the technical side we define the unperturbed
Hamiltonian ${\cal H}_0$ as the $X_A=X_B=\infty$ limit.
In this limit the
two arms are disconnected from each other,
and the diagonalization
gives a set of eigen-energies $E_i$ such that each eigenstate belongs
to only one of the two arms.  Then we make either $X_A$ or $X_B$ or both finite,
and we ask what is the perturbation matrix $W_{ij}$ in the reduced Hamiltonian
\be{3} 
{\cal H}_{ij} 
= \left(\amatrix{E_1 & 0 \cr 0 & E_2}\right)  
+ \left(\amatrix{W_{11} & W_{12} \cr W_{21} & W_{22}}\right)
\ee
Obviously we would like to express the perturbation 
using the transmission coefficients of the barriers.

%%%%%%%%%%%%%%%%%%%%%%%%%%%%%%%%%%%%%%%%%%%%%%%%%%%%%%%%%%%%%%%%%%%%%%%%%%
%%%%%%%%%%%%%%%%%%%%%%%%%%%%%%%%%%%%%%%%%%%%%%%%%%%%%%%%%%%%%%%%%%%%%%%%%%
\section{Outline}
\label{S2}

In the first part of the paper (Sections~\ref{S3}-\ref{S5}) 
we establish the building blocks. 
We derive expressions for the perturbation matrix $W_{ij}$
and for the reduced current operator ${\cal I}_{ij}$,
%and for some other useful operators ${\cal F}_{ij}$,
and figure out how the topology is reflected
in the reduced description.

In the second part of the paper (Sections~\ref{S6}-\ref{S7}) 
we turn to the applications. 
We begin with the simplest problems: 
The coherent Bloch oscillations of a particle
in a mirror symmetric device, 
and the Wigner decay of a particle 
from a short arm to a long arm. 
Then we continue with the quantum stirring problem, 
and show how one can derive expressions 
for the geometric conductance.

In the third part of the paper (Sections~\ref{S8}-\ref{S11}) 
we address some non-trivial technical points 
that are associated with the analysis, 
thus exploring the limitations of the TLS modeling. 
We demonstrate that even if the TLS modeling 
does not apply, we still can use 
a {\em neighboring level approximation} 
in order to extract results for the geometric conductance.

In the Summary (Section~\ref{S12}) we highlight the practical value 
of our findings for the purpose of design and analysis 
of quantum stirring devices, and we briefly 
relate to the experimental measurement issue.

%%%%%%%%%%%%%%%%%%%%%%%%%%%%%%%%%%%%%%%%%%%%%%%%%%%%%%%%%%%%%%%%%%%%%%%%%%
%%%%%%%%%%%%%%%%%%%%%%%%%%%%%%%%%%%%%%%%%%%%%%%%%%%%%%%%%%%%%%%%%%%%%%%%%%
\section{The TLS modeling scheme}
\label{S3}

The unperturbed eigenstates $\psi^{i}(x)$ are labeled 
as ${i=1,2}$, corresponding to the two arms of the ring. 
The associated eigen-energies are ${E_i = k_i^2/(2\mass)}$. 
We have  
\be{4}
\psi^{(1)}(x) =  
\left\{\amatrix{ 
\sqrt{\frac{2}{L_1}}\sin(k_1x+\varphi_1) & \mbox{if $x \in 1$st arm} \cr 
0  &  \mbox{if $x \in 2$nd arm}
} \right.  
\ee
and a similar expression for  $\psi^{(2)}(x)$,  
where $L_i$ is the length of the $i$th arm, 
and $|\varphi|<\pi/2$. 
Two representative eigenstates are illustrated in Fig.2.
Note that the wavenumber of the particle in the $i$th arm 
is $k_i=(\pi/L_i)\times\mbox{\small integer}$. 
Our interest is in a very small 
energy range ${E_1 \sim E_2 \sim E}$, 
where the wavenumbers are ${k_1 \sim k_2 \sim k_E}$, 
corresponding to the velocity ${v_E=(2E/\mass)^{1/2}}$. 
We would like to ignore all the other levels.  
Later we discuss the validity conditions 
for this TLS modeling scheme.

Once we lower from infinity one barrier, say barrier A, 
the two states become coupled. In section \ref{S4} 
we consider a delta barrier and obtain the 
following expression for the perturbation matrix: 
\be{333}
W^A_{ij} = -\frac{v_E}{2\sqrt{L_iL_j}} \ \sqrt{g_A}
\ee
where $g_A$ is the transmission of the barrier.
For $i\neq j$ the minus sign is a convention 
that fixes the gauge (see Appendix~A). 
In section \ref{S5} we show that essentially the
same result applies to any other type of barrier, 
but the ${i=j}$ expression for the energy shift 
should be somewhat generalized.

If both barriers are finite the two associated 
perturbation terms should be added together, 
and one obtains for the energy difference 
\be{6}
\varepsilon = (E_1 + W_{11}^A + W_{11}^B )-(E_2 + W_{22}^A+W_{22}^B)
\ee
and for the coupling 
\be{7}
\frac{\kappa}{2} = W_{12}^A+W_{12}^B = -\frac{v_E}{2\sqrt{L_1L_2}} \left(\sqrt{g_A} \pm\signC \sqrt{g_B}\right)
\ee
The latter expression involves a relative sign $\pm\signC$ 
that cannot be gauged away (see Appendix~A).
If we had magnetic flux penetrating 
through the ring we could have, instead 
of the $\pm\signC$, an arbitrary phase factor. 
  
Using the Pauli matrices we can write the TLS Hamiltonian as 
\be{0}
\mathcal{H}_{ij} \ \ = \ \ 
\frac{\varepsilon}{2} \bm{\sigma}_z + \frac{\kappa}{2} \bm{\sigma}_x  
\ \ \equiv \ \ 
\frac{\bm{\Omega}}{2} \cdot \bf{\sigma} 
\ee
Defining $\theta$ as the angle between
$\Omega$ and the "z" axis, with the convention $0<\theta<\pi$, 
the eigenstates ${n_0}$ and ${m_0}$ of this Hamiltonian are:
\be{9}
|n_0\rangle
=
\left( \begin{array}{c} 
\mp\sin{(\theta/2)} \\ \cos{(\theta/2)}
\end{array} \right),
\ \ \ \  
|m_0\rangle
=
\left( \begin{array}{c} 
\cos{(\theta/2)} \\ \pm\sin{(\theta/2)}
\end{array} \right) 
\ee
where the ${\pm}$ indicates the sign of $\kappa$.
The energy difference between these eigenstates is 
\be{0}
\Omega \ \ = \ \ \sqrt{\varepsilon^2 + \kappa^2}
\ee
If we have a symmetric well then the effective coupling 
between odd and even levels vanishes (${\kappa=0}$), 
and then we can get a degeneracy provided we tune appropriately
the energy level difference~$\varepsilon$.    
  
The TLS description is valid if $W_{12}$ 
is much smaller compared with the level spacing, namely
\be{11}
\mbox{max}\{g_A, \ g_B\} \ \ \ll \ \ L_2/L_1
\ee
where without loss of generality we assume $L_1>L_2$.

In section \ref{S9} we are going to derive expressions 
for the current $\mathcal{I}^A$ through barrier~A, 
as defined by Eq.(\ref{e2}) with ${x_0=x_A}$.  
One observes that the matrix elements of this
operator in the ``standard basis" of Eq.(\ref{e4}) vanish,  
because the unperturbed wavefunctions are zero 
at the barriers. The more careful treatment 
reveals that the reduced operator that gives 
the net current from the first arm to the second arm is 
\be{10}
\mathcal{I}_{ij} \ \ = \ \ \frac{\kappa}{2} \, \bm{\sigma}_y  
\ee
and it turns out that $\mathcal{I}_{ij}^A=\lambda_A\mathcal{I}_{ij}$,  
and $\mathcal{I}_{ij}^B=\lambda_B\mathcal{I}_{ij}$, 
where the splitting ratio is defined as 
\be{12}
\lambda_A=\frac{W_{12}^A}{W_{12}^A+ W_{12}^B}=
\frac{\sqrt{g_A}}{\sqrt{g_A}\pm\signC\sqrt{g_B}}
\ee
with a similar definition for $\lambda_B$.
We have ${\lambda_A+\lambda_B=1}$, but contrary 
to the naive point of view ${0<\lambda_A<1}$ is not implied.
Rather, if the two states have opposite parity, 
then one~$\lambda$ is larger than $100\%$, 
while the other~$\lambda$ is negative. 
We shall see later in Section \ref{S12} that the physical 
interpretation of the ``splitting ratio" 
requires recognition in the existence of induced 
circulating current in the system.
Thus the multiple path topology of the system 
is reflected in the TLS modeling via~$\lambda$.

%%%%%%%%%%%%%%%%%%%%%%%%%%%%%%%%%%%%%%%%%%%%%%%%%%%%%%%%%%%%%%%%
%%%%%%%%%%%%%%%%%%%%%%%%%%%%%%%%%%%%%%%%%%%%%%%%%%%%%%%%%%%%%%%%
\section{The expression for $W_{ij}$ for a delta barrier}
\label{S4}

Let us assume that barrier~B is infinitely high, 
while barrier~A is modeled as a delta function. 
In other words: we consider the simplest possibility 
of having an infinite well [${(-L/2)<x<(L/2)}$] 
which is divided at ${x=x_A}$ by a delta function:
\be{0}
V_A(x-x_A) \ \ = \ \ X_A\delta(x-x_A)
\ee
The total perturbation is obtained from  
a sequence of infinitesimal variations 
of the barrier height
\be{13}
{\cal H}(X_A) 
\ \ &=& \ \ 
{\cal H}(\infty)-\int^\infty_{X_A}
\frac{\partial{\cal H}}{\partial X} \ \mbox{d}X 
\\
\ \ &\equiv& \ \ {\cal H}(\infty)+W^A
\ee
For any value of $X$ the Hilbert space of the system 
is spanned by a set of (real) eigenfunction labeled by~$n$. 
The matrix elements for an infinitesimal variation 
of the barrier height is 
\be{0}
\left(\frac{\partial{\cal H}}{\partial X}\right)_{nm}
\ \ = \ \ \psi^{(n)}(x_A) \,\, \psi^{(m)}(x_A)
\ee
Using the matching conditions for a delta potential 
at ${x=x_A}$ we can express the wave function by its
derivative:
\be{0}
\psi^{(n)}(x_A) 
= \frac{1}{2\mass X_A} \left[\partial\psi^{(n)}(x_A{+}0)-\partial\psi^{(n)}(x_A{-}0)\right]
\ee
A more elegant way of writing this relation is  
\be{16}
\psi^{(n)}(x_A) 
= \frac{1}{2\mass X_A} \sum_{a=1,2} \partial_a\psi^{(n)}(x_A)
\ee
where $\partial_a$ is defined as the {\em radial} derivative 
in the direction of the $a$th arms that stretch out 
of the junction at ${x=x_A}$. Defining the total 
radial derivative as ${\partial=\partial_1+\partial_2}$
we get 
\be{0}
\left(\frac{\partial{\cal H}}{\partial X}\right)_{nm}
=
\frac{1}{(2\mass X_A)^2}
\,\,
\partial\psi^{(n)}(x_A)
\,\,
\partial\psi^{(m)}(x_A)
\ee
For a large barrier with small transmission 
\be{18}
g_A \ \ \approx \ \ \left(\frac{v_E}{X_A}\right)^2 \ \ \ll \ \ 1 
\ee
the $n$th and $m$th states remain similar 
to some unperturbed $i$th and $j$th states. 
Accordingly, upon integration we get 
from Eq.(\ref{e13}) the result 
\be{20}
W^A_{ij} = -\frac{1}{4\mass^2 X_A}
\Big[\partial\psi^{(i)}(x_A)\Big]
\,\,
\Big[\partial\psi^{(j)}(x_A)\Big] 
\ee
Note that in the last equation the contribution to the 
total derivative $\partial$ comes from one term only, 
because each unperturbed wavefunction $\psi^{(i)}(x)$ is non-zero 
only in one box. Using  Eq.(\ref{e18}) we get Eq.(\ref{e333}).

%%%%%%%%%%%%%%%%%%%%%%%%%%%%%%%%%%%%%%%%%%%%%%%%%%%%%%%%%%%%%%%%%%%%
%%%%%%%%%%%%%%%%%%%%%%%%%%%%%%%%%%%%%%%%%%%%%%%%%%%%%%%%%%%%%%%%%%%%
\section{The expression for $W_{ij}$ for a general barrier} 
\label{S5}

It is possible to deduce an expression for $W_{ij}$
without assuming a specific form of potential barrier.
For the purpose of this calculation we describe the barrier 
at ${x=x_A}$ by a general scattering matrix 
\be{0}
S &=& \eexp{i\gamma}
\left(
\begin{array}{cc}
i\sqrt{1-g} \ \eexp{i\alpha} & -\sqrt{g} \\
-\sqrt{g} & i\sqrt{1-g} \ \eexp{-i\alpha}
\end{array}
\right)
\ee
Regarding the barrier as a {\em junction} it can be 
embedded either in a {\em closed} ring geometry with the two {\em arms} attached,  
or in an {\em open} one-dimensional geometry with two infinite {\em leads} attached.
In both cases the differential representation 
of $W$ should be the {\em same}, because $W$ is local in space. 
In other words $W^A_{ij}$ should come out the same 
for the wavefunctions $\psi^{(i)}(x)$ and $\psi^{(j)}(x)$
of the ring, if in the vicinity of ${x=x_A}$ 
they are identical with $\Psi^{(i)}(x)$ and $\Psi^{(j)}(x)$  
of the scattering geometry.

In the scattering geometry it is conventional 
to label the two leads by ${a=1,2}$ 
and to define a radial coordinate $r=|x-x_A|$.
The flux normalized scattering states 
of the junction (assuming outgoing waves) 
are $\Psi^{(i+)}$. By definition we have   
\be{0}
\Psi^{(1+)} = 
\left\{\amatrix{ 
\frac{1}{\sqrt{v_E}} [\eexp{-ik_E r} - S_{11}\eexp{ik_E r}] & \mbox{if $r \in 1$st lead} \cr 
\frac{1}{\sqrt{v_E}} [-S_{21}\eexp{ik_E r}] &  \mbox{if $r \in 2$nd lead}
} \right.  
\ee
A similar expression holds for $\Psi^{(2+)}$.
If the leads are not coupled, the scattering 
matrix becomes $S_0$ with ${g=0}$.  
In the vicinity of ${x=x_A}$ the unperturbed scattering states  
coincide with those of Eq.(\ref{e4}) up to normalization. 
Namely, in the vicinity of ${x=x_A}$ we have the relation  
\be{4444}
\Psi^{(i)}(x) = -i\left(\frac{2L_i}{v_E}\right)^{1/2} 
\eexp{i\varphi_i} 
\,\, \psi^{(i)}(x)
\ee
where 
\be{0}
\varphi_i = \frac{1}{2} \left( \gamma_0 + \frac{\pi}{2} \pm \alpha_0 \right)
\ee
with $\pm$ sign for $i=1,2$ respectively.

The relation between the scattering matrix
and the perturbation matrix~$W$ can be 
deduced via the $T$~matrix formalism. 
The $S$ matrix is related to the $T$ matrix 
through ${S = (1-iT)S_0}$, or more explicitly  
\be{0}
[SS_0^{-1}]_{ij} = \delta_{ij} - i\langle \Psi^{(i)} | T | \Psi^{(j)} \rangle
\ee
In leading order $T$ equals $W$ so we have  
\be{0}
\langle \Psi^{(i)} | W | \Psi^{(j)} \rangle \ \ \approx \ \ i (S-S_0) \ S_0^{-1}
\ee
where
\be{0}
S-S_0 =
\eexp{i\gamma_0}
\left(
\begin{array}{cc}
\eexp{i\alpha_0}(\delta\gamma+\delta\alpha) & \sqrt{g} \\
\sqrt{g} & \eexp{-i\alpha_0}(\delta\gamma-\delta\alpha)
\end{array}
\right)
\ee
Thus
\be{0}
\langle \Psi^{(i)} | W | \Psi^{(j)} \rangle  = 
-\left(
\begin{array}{cc}
\delta\gamma+\delta\alpha & \sqrt{g} \ \eexp{i\alpha_0} \\
\sqrt{g} \ \eexp{-i\alpha_0} & \delta\gamma-\delta\alpha
\end{array}
\right)
\ee
Using Eq.(\ref{e4444}) we deduce that each 
element of ${\langle \psi^{(i)} | W | \psi^{(j)} \rangle}$ 
involves multiplication by ${v_E/(4 L_i L_j)^{1/2}}$, 
while the $\alpha_0$ is canceled out. 
This leads to Eq.(\ref{e333}) for the ${i \ne j}$ coupling, 
and a generalized expression for the energy level shifts.

%%%%%%%%%%%%%%%%%%%%%%%%%%%%%%%%%%%%%%%%%%%%%%%%%%%%%%%%%%%%%%%%%%%%%%%%%%%%
%%%%%%%%%%%%%%%%%%%%%%%%%%%%%%%%%%%%%%%%%%%%%%%%%%%%%%%%%%%%%%%%%%%%%%%%%%%%
\section{Wigner decay and Bloch oscillations}
\label{S6}

If the two arms of the ring have exactly the same length ${L_1=L_2=L/2}$,
then the coherent Bloch oscillations of a wavepacket 
in such a symmetric double well are characterized by the frequency
\be{0}
\Omega_{\tbox{Bloch}} = 2|W_{12}| = \frac{v_E}{L_1}\left|\sqrt{g_A}+\sqrt{g_B}\right|
\ee

If one arm of the ring ($L_1$) is very long,  
and the other arm ($L_2$) is short,
then a particle placed initially at the short arm
will decay into the quasi continuum of the long arm.
The decay rate is given by the Fermi golden rule
\be{27}
\Gamma = \frac{2\pi}{\Delta}|W_{12}|^2
\ee
where $\Delta=(\pi/L)v_E$ is the mean level spacing. 
If the arms are coupled through barrier~A,  
while barrier~B is infinitely high, 
then the decay rate is
\be{0}
\Gamma &=& \frac{v_E}{2L_2} \ g_A 
\ee
This result agrees with the well known Gamow formula: 
the decay rate is given by the attempt
frequency multiplied by the probability 
to cross the barrier.

If both barriers are finite, it is important to notice that 
the quasi-continuum of the long arm is composed of odd and even states.
The state of the short arm, which is either even or odd, is coupled to states of
the same parity with a plus sign in the expression of Eq.(\ref{e7}) 
and to states of the opposite parity with a minus sign.
Accordingly, the decay rate is the sum of the decay rate to states of
the same parity and the decay rate to states of the opposite parity
\be{0}
\Gamma 
\ \ &=& \ \ 
\sum_{\pm}
\frac{2\pi}{\Delta_{\pm}}
\left|
\frac{v_E}{2\sqrt{L_1L_2}}
\left(\sqrt{g_A} \pm\signC \sqrt{g_B}\right)
\right|^2
\nonumber\\
\ \ &=& \ \
\frac{v_E}{2L_2}\left(g_A+g_B\right)
\ee
where $\Delta_{\pm}=2\Delta$ is the mean level spacing 
for states with the same parity. So inspite of the parity 
considerations we still get the naive result that agree 
with Gamow formula.

%%%%%%%%%%%%%%%%%%%%%%%%%%%%%%%%%%%%%%%%%%%%%%%%%%%%%%%%%%%%%%%%%%%%%%%%%%%%
%%%%%%%%%%%%%%%%%%%%%%%%%%%%%%%%%%%%%%%%%%%%%%%%%%%%%%%%%%%%%%%%%%%%%%%%%%%%
\section{Quantum stirring}
\label{S7}

We assume that we have control over geometrical parameters
of the device, such as the potential floor in each arm, 
the barriers heights, their location, or any other gate controlled 
feature of the potential landscape. With a control parameter~$X$ 
we associate a generalized force operator 
\be{21}
{\cal F} = -\frac{\partial{\cal H}}{\partial X}
\ee
Quantum stirring means to induce 
a circulating current by changing 
the parameter~$X$. We assume that the parametric variation is 
adiabatic so we have a linear relation $\langle I \rangle = - G \dot{X}$,
where $G$ is know as the geometric conductance \citeref{pmc}.
The Kubo formalism implies that $G$ equals to the Berry curvature \citeref{berry1,avron2,berry2}:
\be{22}
G = \sum_{m(\neq n)}\frac{2 \ \im[{\cal I}_{nm}]{\cal F}_{mn}}{(E_m-E_n)^2}
\ee
where $n$ is the level in which the particle is prepared, 
and $m$ are the other levels.

Within the framework of the TLS modeling the sum in Eq.(\ref{e22}) 
contains only one term which involves the states $n_0$ and $m_0$ 
of Eq.(\ref{e9}). For the matrix element of the current operator we get
\be{34}
{\cal I}_{n_0m_0} 
= \left[\lambda\frac{\kappa}{2}\bm{\sigma}_y\right]_{n_0m_0}  
= i\lambda \, \frac{\kappa}{2} 
\ee
where $\lambda$ is the appropriate splitting ratio. 
The matrix element of the generalized force 
operator is calculated using Eqs.(\ref{e9}) and (\ref{e21}) 
\be{35}
{\cal F}_{m_0n_0} 
&=& 
-\frac{1}{2} 
\left[
\frac{\partial \varepsilon}{\partial X}\bm{\sigma}_z 
+ \frac{\partial \kappa}{\partial X} \bm{\sigma}_x
\right]_{m_0n_0} 
\\
&=& 
\pm \ \frac{1}{2}\sin(\theta) \ \frac{\partial \varepsilon}{\partial X} 
-\frac{1}{2}\cos(\theta) \ \frac{\partial \kappa}{\partial X} 
\\
&=& 
\frac{1}{2\Omega}
\left(
\kappa \frac{\partial \varepsilon}{\partial X} 
- \varepsilon \frac{\partial \kappa}{\partial X} 
\right)
\ee
where we used ${\pm\sin(\theta)=\kappa/\Omega}$
and ${\cos(\theta)=\varepsilon/\Omega}$. 
This leads for the following result for the 
geometric conductance: 
\be{36}
G =   
\frac{\lambda\kappa}{2\Omega^3} 
\left[ 
\kappa \frac{\partial \varepsilon}{\partial X} 
- \varepsilon \frac{\partial \kappa}{\partial X} 
\right]
\ee
In the analysis of the operation of a stirring device 
we typically have a well defined region  
where the potential is being varied. We may call 
this segment ``the pump". It is convenient to measure 
the current elsewhere, where the potential is fixed.  
If barrier~A is not part of the ``pump"
then we can measure the current at $x_0=x_A$. 
Then it follows from the definitions of $\lambda$ 
and $\kappa$ that the product $\lambda\kappa$ 
does not change with time, even if barrier~B is modulated.
Then we can rewrite the above formula as 
\be{0}
G =   
\frac{\lambda_0\kappa_0}{2\Omega^3} 
\left[ 
\kappa \frac{\partial \varepsilon}{\partial X} 
- \varepsilon \frac{\partial \kappa}{\partial X} 
\right]
\ee
where $\lambda_0$ and $\kappa_0$ are that values at 
some arbitrary moment of time.  
Typically the variation of $X$ leads to a sequence 
of level crossings if $\kappa$ is disregarded.
These become avoided crossings if $\kappa$ 
is taken into account.
At the vicinity of a crossing we typically 
can use a linear approximation:
\be{0} 
\varepsilon 
\ \ &=& \ \ \dot{\varepsilon}  \times (X-X_0)
\\
\kappa 
\ \ &=& \ \ \kappa_0 + \dot{\kappa} \times  (X-X_0)
\ee
The amount of probability $dQ=Idt$ which is being 
transported equals ${-GdX}$. For an individual crossing  
the $dX$ integration over $G$ can be performed using 
\be{0}
&&\int_{-\infty}^{+\infty}\frac{a(b+cx) \ \mbox{d}x}
{\left(a^2x^2+(b+cx)^2\right)^{3/2}}=\frac{2a}{b\sqrt{a^2+c^2}}\\
&&\int_{-\infty}^{+\infty}\frac{c \, ax \ \mbox{d}x}
{\left(a^2x^2+(b+cx)^2\right)^{3/2}}=-\frac{2c^2}{ab\sqrt{a^2+c^2}}
\ee
Then we get the result
\be{44}
Q \ \ = \ \ \pm \lambda_0 \sqrt{1+\left({\dot{\kappa}}/{\dot{\varepsilon}}\right)^2}
\ee
where the $\pm$ is determined according to the sign 
of~$\dot{\varepsilon}$. We observed that in order to get 
the ``quantized" value ${Q=1}$ there should be neither topological  
splitting (${\lambda=1}$) nor barrier modulation (${\dot{\kappa}=0}$) 
during the transition.

%%%%%%%%%%%%%%%%%%%%%%%%%%%%%%%%%%%%%%%%%%%%%%%%%%%%%%%%%%%%%%%%%%%%%%%%%%%%
%%%%%%%%%%%%%%%%%%%%%%%%%%%%%%%%%%%%%%%%%%%%%%%%%%%%%%%%%%%%%%%%%%%%%%%%%%%%
\section{The neighboring level approximation scheme}
\label{S8}

A major interest is in systems with zero temperature Fermi occupation. 
In such a case Eq.(\ref{e22}) has to be summed over $n$ up to 
the Fermi energy. It turns out \citeref{pmx} that the result is dominated 
by the contribution that come from the coupling between the last 
occupied level and its neighboring empty level. This suggests 
to adopt a neighboring level approximation scheme that holds 
irrespective of the validity of the TLS modeling, and coincides  
with it if the condition of Eq.(\ref{e11}) is satisfied. 
The key idea is to characterize each eigenstate by a mixing parameter 
\be{0}
\Theta \ \ \equiv \ \ 2\arctan\left(\sqrt{\frac{\mbox{Prob}(x\in2)}{\mbox{Prob}(x\in1)}}\right) 
\ee
such that $\Theta=0$ for states that belong to the first arm 
and $\Theta=\pi$ for states that belong to the second arm.
Numerical examples are presented in Figs.~3-4. 
If we are given $\Theta$ then we can construct the eigenstate 
using a procedure that we describe below. If the TLS modeling 
applies then $\Theta$ becomes essentially the same as $\theta$.

Let us see how we construct the wavefunction given the 
energy ${E=E_n}$, the mixing parameter ${\Theta=\Theta_n}$, 
and the parity $\pm\signA$ with respect to (say) barrier~A,
as defined in Appendix~A. Consequently it is convenient 
to set the origin such that ${x_A=0}$, 
and write the $n$th eigenstate of the ring as 
\be{37}
\psi^{(n)}(x) =  
\left\{\amatrix{ 
\pm\signA C_1\sin(k_1x+\varphi_1) & \mbox{if $x \in 1$st arm} \cr 
C_2\sin(k_2x+\varphi_2) & \mbox{if $x \in 2$nd arm}
} \right.  
\ee
where $C_i>0$, and $|\varphi|<\pi/2$. 
Assuming $k_EL\gg1$, the amplitudes satisfy the normalization condition 
\be{0}
\frac{1}{2}L_1{C_1}^2+\frac{1}{2}L_2{C_2}^2 \approx 1
\ee
It follows that 
\be{40}
C_1 &\approx& \sqrt{\frac{2}{L_1}} \ \cos\left(\frac{\Theta}{2}\right)\\
\label{e41}
C_2 &\approx& \sqrt{\frac{2}{L_2}} \ \sin\left(\frac{\Theta}{2}\right)
\ee
We still have to say what are 
the wavenumbers $k_1$ and $k_2$,   
and the phase shifts $\varphi$. 
Let us see first how they are determined 
within the framework of the TLS modeling, 
and then how they can be found irrespective 
of the TLS modeling.

Naively the $|n_0\rangle$ and $|m_0\rangle$ eigenstates, 
within the framework of the TLS modeling, 
are the superposition of the basis states 
of Eq.(\ref{e4}) and accordingly  
\be{42}
\Theta &=& \theta, \, \pi{-}\theta \\
\label{e443}
k_1 &=& \mbox{corresponds to the unperturbed $E_{1}$} \\
\label{e444}
k_2 &=& \mbox{corresponds to the unperturbed $E_{2}$} \\
\varphi_1 &=& \mbox{same as the unperturbed}  \\
\varphi_2 &=& \mbox{same as the unperturbed} 
\ee
while the true eigenstates are with (see Fig.2)
\be{47}
\Theta &\approx& \theta, \, \pi{-}\theta \\
k_1 &=& \mbox{corresponds to $E_n$} \\
k_2 &=& \mbox{corresponds to $E_n$} \\
\varphi_1 &=& \mbox{shifted}  \\
\varphi_2 &=& \mbox{shifted}
\ee
To be more specific, we have $\Theta^{(m_0)} \approx \theta$
and $\Theta^{(n_0)} \approx \pi-\theta$ 
and hence ${\Theta^{(m_0)}{+}\Theta^{(n_0)}\approx \pi}$ 
if the TLS modeling is valid (see Fig.~4). 
We note that from Eq.(\ref{e40}-\ref{e41}) it follows 
that within the framework of the TLS approximation we have   
\be{52}
C_i^{(m_0)}C_i^{(n_0)} \ \ \approx \ \ \frac{1}{L_i} \sin(\theta) 
\ee
This will be used later on in order to obtain 
simplified expressions for the matrix elements 
of various operators.

Whether $k_1$ and $k_2$ in Eqs.(\ref{e443}-\ref{e444}) correspond to the 
same energy or not, is not a big difference 
for us because we assume ${k_1 \sim k_2 \sim k_E}$ 
in any case. The main problem with the naive version 
is related to the phase shifts, as demonstrated in Fig.~2.
The variation of the phase shift as 
the barriers are lowered reflect that there 
is a non-zero probability to find the particle 
in the region of the barriers.  
In particular if the phases $\varphi_i$ remained 
the same it would imply that all the matrix 
elements of $I^A$ and $I^B$ would be zero.  
It is essential to take the variation of  $\varphi$
into account in order to get a non-zero result 
for the geometrical conductance. 
We shall discuss the calculation of the matrix 
elements $\mathcal{I}_{nm}$ and $\mathcal{F}_{nm}$     
in the next sections. First we would like to discuss 
how the required information on the variation 
of the phases~$\varphi$ can be extracted.

In order to express $\varphi$ by $\Theta$,
we write the wave function of Eq.(\ref{e37})
as ingoing and outgoing waves and set the origin $x=0$ at
either one of the barriers, for example barrier A.
We match the wave functions of the two bonds by
the barrier scattering matrix
\be{58}
\left(
\begin{array}{c}
{\pm\signA} C_1 \ \eexp{+i\varphi_1} \\
C_2 \ \eexp{+i\varphi_2}
\end{array}
\right)
 &=& {\bm S}_A
\left(
\begin{array}{c}
{\pm\signA} C_1 \ \eexp{-i\varphi_1} \\
C_2 \ \eexp{-i\varphi_2}
\end{array}
\right)
\ee
and get closed equations for the phase shifts
\be{48}
\sqrt{1-g} \sin(2\varphi_1-\alpha-\gamma) {=}
1-\frac{g}{2}\left(1+\left(\frac{C_2}{C_1}\right)^2\right)\\
\sqrt{1-g} \sin(2\varphi_2+\alpha-\gamma) {=}
1-\frac{g}{2}\left(1+\left(\frac{C_1}{C_2}\right)^2\right)
\ee

So far everything is exact. So once we have $\Theta$
we can find the phases and construct the wavefunction.
We would like to focus in the rest of this section 
in the regime where the TLS modeling applies. 
Assuming that $\Theta$ is determined by $\theta$ 
we want to find what are $\varphi_1$ and $\varphi_2$, 
so as to construct a proper wavefunction. 
Neglecting terms of order $g$ and expanding $\arcsin(1{-}x)$ 
as $\pi/2\pm\sqrt{2x}$ we obtain  
\be{49}
\varphi_1 &\approx& \frac{\gamma+\alpha}{2}+\frac{\pi}{4}
\pm\signA\frac{\sqrt{g}}{2} \ \sqrt{\frac{L_1}{L_2}} \ 
\tan\left(\frac{\Theta}{2}\right)
\ee
where the $\pm\signA$ sign should be the same as in Eq.(\ref{e58}), 
which can be established by direct substitution. 
A similar expression can be obtained for $\varphi_2$.
We note that within the framework of this TLS approximation we have
\be{53}
\varphi_1^{(m_0)}-\varphi_1^{(n_0)} =
\pm\sqrt{g_A} \left(\frac{L_1}{L_2}\right)^{1/2} \frac{1}{\sin(\theta)}
\ee
where the sign is the same as that of $\kappa$.
We now have all the building blocks needed
for the calculation of the matrix elements.

%%%%%%%%%%%%%%%%%%%%%%%%%%%%%%%%%%%%%%%%%%%%%%%%%%%%%%%%%%%%%%%%
%%%%%%%%%%%%%%%%%%%%%%%%%%%%%%%%%%%%%%%%%%%%%%%%%%%%%%%%%%%%%%%%
\section{The expression for $\mathcal{I}_{nm}$}
\label{S9}

If we adopt the TLS point of view,
we can postulate a self-consistent definition 
of the current operator based on the continuity equation. 
For this purpose we define the occupation operator ${\cal N}$ 
for one of the arms as
\be{0}
{\cal N} 
\ \ = \ \ 
\left(
\begin{array}{cc}
1 & 0   \\
0 & 0
\end{array}
\right)
\ee
and deduce the definition of the current operator from 
\be{0}
\frac{d}{dt}{\cal N} 
\ \ = \ \  
i[\mathcal{H},\mathcal{N}]  
\ \ \equiv \ \ {\cal I}
\ee
where ${\cal I}$ is given by Eq.(\ref{e10}).
If we turn off the coupling at barrier~A 
we get the same expression multiplied by $\lambda_B$, 
while if we turn off the coupling at barrier~B 
we get the same expression multiplied by $\lambda_A$.  

The above reasoning bypass the confrontation which is 
involved in carrying out a direct calculation, 
and hence contains an uncontrolled error 
which is associated with the assumption that a TLS  
description of Hilbert space is valid. 
If we revert to the original definition of Eq.(\ref{e2}),
then the matrix elements are given by
\be{0}
{\cal I}_{nm} = i\frac{1}{2\mass}\left(
\partial\psi^{(n)} \ \psi^{(m)}-\psi^{(n)} \ \partial\psi^{(m)}\right)
\Big|_{x=x_0}
\ee
For the calculation of ${\cal I}^A_{nm}$ we set $x_0=x_A=0$.
As was already pointed out, in order to get a non-trivial 
result, we have to take into account the phase shifts $\varphi$
which was calculated in the previous section.
Substituting the wave function of Eq.(\ref{e37}) we get
\be{61}
{\cal I}^A_{nm} &=& 
-i\frac{1}{2\mass}C_1^{(m)}C_1^{(n)}\left[
\frac{k_{m}+k_{n}}{2} \ \sin(\varphi_1^{(m)}-\varphi_1^{(n)})
\right.\nonumber\\
&&+ \left.\frac{k_{n}-k_{m}}{2} \ \sin(\varphi_1^{(m)}+\varphi_1^{(n)})
\right]
\ee
Whenever the TLS modeling applies we can substitute
Eqs.(\ref{e52}) and (\ref{e53}) into Eq.(\ref{e61}). 
Neglecting the second term we get
\be{70}
{\cal I}^A_{n_0m_0} \approx {\mp}i\frac{v_{\tbox{E}}}{2\sqrt{L_1L_2}}\sqrt{g_A}
\ee
where the sign is the same as that of $-\kappa$.
One notices that the expression for ${\cal I}^A_{n_0m_0}$ 
can be written as Eq.(\ref{e34}) where $\kappa$ 
and $\lambda$ are given by Eq.(\ref{e7}) and Eq.(\ref{e12}).

%%%%%%%%%%%%%%%%%%%%%%%%%%%%%%%%%%%%%%%%%%%%%%%%%%%%%%%%%%%%%%%%%%%%
%%%%%%%%%%%%%%%%%%%%%%%%%%%%%%%%%%%%%%%%%%%%%%%%%%%%%%%%%%%%%%%%%%%%
\section{Stirring by barrier modulation}
\label{S11}

In this section we calculate the geometric conductance 
as determined by the matrix elements of the generalized force 
that is associated with modulation
of a delta barrier.  
The motivation is to verify the results of the reduced 
description against the direct full Hilbert space calculation. 
The potential barrier is given by
\be{0}
V_B(\hat{x}) = X_B\delta(\hat{x}-x_B)
\ee
The stirring is induced by variation of the barrier
height $X_B$. The associated generalized force is 
\be{0}
{\cal F} = -\frac{\partial{\cal H}}{\partial X_B} 
\ \ = \ \ -\delta(\hat{x}-x_B)
\ee
with the matrix elements 
\be{0}
{\cal F}_{mn} \ \ = \ \ -\psi^{(n)}\psi^{(m)}
\ee
For the wavefunctions amplitudes we use Eq.(\ref{e52})
and for the phase shifts Eq.(\ref{e49}).
We also substitute the scattering matrix parameters
that describe a delta barrier
\be{0}
\gamma_B &\approx& -\pi/2+\sqrt{g_B}\\
\alpha_B &=& 0
\ee 
where the approximation is valid for $g_B\ll1$ and
the relation of $g_B$ and $X_B$ is given in Eq.(\ref{e18}).
With the above approximations we get 
\be{0}
{\cal F}_{m_0n_0} \approx g_B\frac{L_2-L_1}{4L_1L_2}\sin(\theta) \ {\mp} \
\frac{g_B}{2\sqrt{L_1L_2}}\cos(\theta)
\ee 
where the sign should be the same as that of $\mp\signC\kappa$. 
In order to verify the consistency with the TLS expression, 
we differentiate Eq.(\ref{e6}) and Eq.(\ref{e7}): 
\be{0}
\frac{\partial \varepsilon}{\partial X_B} &=&
g_B\frac{L_2-L_1}{2L_1L_2}\\  
\frac{\partial \kappa}{\partial X_B} &=&
\pm\signC\frac{g_B}{\sqrt{L_1L_2}}
\ee
and substitute into Eq.(\ref{e35}).
Indeed we obtain the same result 
for ${\cal F}_{m_0n_0}$ as above.

The geometric conductance of Eq.(\ref{e22}) involves
the multiplication of ${\cal F}_{m_0n_0}$ with ${\cal I}_{n_0m_0}$, leading to
\be{0}
G &=& \frac{1}{4} \ v_{E}^2 \frac{L_2-L_1}
{\left(L_1L_2\right)^2} \ \frac{g_A \ g_B\pm\signC{g_A}^{1/2}{g_B}^{3/2}}{\Omega^3}
\nonumber\\
&\mp\signC& \frac{1}{4} \ v_{E}^2 \frac{L_2+L_1}
{\left(L_1L_2\right)^2} \ \frac{g_A \ g_B+{g_A}^{1/2}{g_B}^{3/2}}{\Omega^3}
\ee 
The calculation of the transport proceeds as in Section~\ref{S7}.

%%%%%%%%%%%%%%%%%%%%%%%%%%%%%%%%%%%%%%%%%%%%%%%%%%%%%%%%%%%%%%%%%%%%%%%
%%%%%%%%%%%%%%%%%%%%%%%%%%%%%%%%%%%%%%%%%%%%%%%%%%%%%%%%%%%%%%%%%%%%%%%
\section{Stirring by barrier translation}
\label{S10}

In complete analogy with the previous section we  
would like to calculate the geometric conductance
as determined by the matrix elements  
of the generalized force which is associated 
with the translation of the barrier: 
\be{0}
{\cal F} = -\frac{\partial{\cal H}}{\partial x_B} \ = \
X_B\delta'(\hat{x}-x_B)
\ee
One obtains 
\be{0}
{\cal F}_{mn} = 
-X_B\left(\overline{\partial\psi^{(n)}} \ \psi^{(m)}+
\overline{\partial\psi^{(m)}} \ \psi^{(n)}\right)
\ee
where $\overline{\partial\psi}$ is the
average derivative on both sides of the barrier.
We simplify this expression by using Eq.(\ref{e16}):
\be{0}
{\cal F}_{mn}=\frac{1}{2\mass}
\left[\partial\psi_1^{(n)}\partial\psi_1^{(m)}-
\partial\psi_2^{(n)}\partial\psi_2^{(m)}
\right]_{x=x_B}
\ee
Assuming high barriers we get in leading order 
\be{0}
{\cal F}_{mn} \approx -\frac{1}{2}\mass v_{E}^2\left( C_1^{(m)}C_1^{(n)} + C_2^{(m)}C_2^{(n)}\right)
\ee
which together with Eq.(\ref{e52}) leads to 
\be{0}
{\cal F}_{m_0n_0} \approx -\frac{1}{2}\mass v_{E}^2 \ \frac{L_1+L_2}{L_1L_2} \ \sin(\theta)
\ee 
In order to compare the above result for ${\cal F}_{m_0n_0}$ 
with the TLS result of Eq.\ref{e35}
we calculate the variation of the potential floor
by taking in Eq.(\ref{e6}) the energies of infinite wells
with $L_1=x_B$ and $L_2=L-x_B$. We get
\be{0}
\frac{\partial \varepsilon}{\partial x_B} &=& -\mass v_E^2
\ \frac{L_1+L_2}{L_1L_2} +\mathcal{O}(\sqrt{g})\\
\frac{\partial \kappa}{\partial x_B} &=& v_E
\ \frac{L_2-L_1}{\left(L_1L_2\right)^{3/2}}
\left(\sqrt{g_A}\pm\signC\sqrt{g_B}\right)
\ee
Substitute into Eq.(\ref{e35}) indeed leads 
to the same result for ${\cal F}_{m_0n_0}$ as above.
Note that in this case (unlike the previous section) 
the second term in Eq.(\ref{e35}) which involves 
the variation of $\kappa$ is of higher order in $g_B$
and therefore should be excluded.

The geometric conductance of Eq.(\ref{e22}) involves
the multiplication of ${\cal F}_{m_0n_0}$ with ${\cal I}_{n_0m_0}$, leading to
\be{0}
G \ \ = \ \  -\frac{1}{2}\sqrt{g_{A}} \ \mass v_{E}^4 \
\frac{L_1+L_2}{\left(L_1L_2\right)^2} \ \frac{\sqrt{g_A}\pm\signC\sqrt{g_B}}{\Omega^3}
\ee 
The calculation of the transport proceeds as in Section~\ref{S7}.  
One realizes that a translation of the barrier 
is effectively equivalent to the variation of the potential floor difference,   
as long as it does not involve modulation of its transmission 
(which is assumed to be small).

%%%%%%%%%%%%%%%%%%%%%%%%%%%%%%%%%%%%%%%%%%%%%%%%%%%%%%%%%%%%%%%%%%%%%%%
%%%%%%%%%%%%%%%%%%%%%%%%%%%%%%%%%%%%%%%%%%%%%%%%%%%%%%%%%%%%%%%%%%%%%%%
\section{Error estimates and limitations}
\label{S13}

If we vary a parameter $X$ then the energy levels $E_n(X)$ form 
a ``spaghetti" which is characterized by a mean level spacing~$\Delta$ 
and possibly by narrow avoided crossings with splitting~$\Delta_0$. 
For the ring system that we are considering 
it follows from the estimate of $\kappa$ that
\be{0}
\frac{\Delta_0}{\Delta} \ \ \sim \ \ \mbox{min}\{1, \sqrt{bg}\}
\ee 
where $b=L_1/L_2$ and $g=\mbox{max}\{g_A,g_B\}$.
The condition Eq.(\ref{e11}) for the applicability 
of the TLS modeling ensures $\Delta_0\ll\Delta$.
In such circumstances Eq.(\ref{e22}) for the geometric 
conductance, which in essence is a sum 
of the type ${\sum_{n=0}^{\infty} (\Delta_0+ n\Delta)^{-2}}$, 
implies that the error that is involved 
in the neighboring level approximation is 
\be{0}
\frac{\mbox{error}(G)}{G} 
\ \ \sim  \ \  \left( \frac{\Delta_0}{\Delta} \right)^2  
\ \  \sim  \ \ bg \ \  \ll 1
\ee 
Once the TLS modeling fails the error becomes of order unity. 
This sounds bad, but in fact it is not so bad. 
The good news is that the far levels contribute to~$G$
a correction which is of the same order as the leading term. 
Therefore with the neighboring level approximation 
we can still get a realistic estimate disregarding numerical  
prefactors of order unity.

Having $b\gg1$ is very interesting, because then we have 
a non-trivial intermediate regime ${1/b \ll g \ll 1}$  
where neither 1st order perturbation theory with respect 
to ``zero" height barriers, nor 1st order perturbation 
theory with respect to ``infinite" barriers applies. 
This is the regime where each level of the small arm 
forms a distinct Wigner resonance with the 
quasi-continuum states of the long arm. 
Obviously the TLS modeling is not applicable in this regime,  
but the neighboring level approximation still provides 
a decent starting  point for a calculation. We shall  
explore this Wigner regime in a future work.

One may also wonder whether the specific results that we have 
obtained for stirring using a {\em delta} barrier applies also 
for a {\em thick} barrier. On physical grounds it is quite 
obvious that the induced current is determined by  
the scattering matrix of the modulated barrier.  
Consequently if the $S(E)$ of the modulated barrier is $E$~independent 
within the energy range of interest, it can be regarded 
as representing a delta function, and the results should come out the same.

Finally one may wonder about the implications of finite 
temperature or non-adiabatic driving. These aspects are 
complementary to the theme of the present paper.
Namely, as discussed in \citeref{pmx}, at finite temperatures 
the statistics of the occupation should be taken into account. 
So we have to average (so to say) over the level that we have  
labeled as~$n_0$ with an appropriate weight as implied by 
the Fermi function. On the other hand the non-adiabatic effects 
require to introduce in the denominator of the 
Kubo formula Eq.(\ref{e22}) a term that represents 
the ``width" of the Fermi-golden-rule transitions.    
Then the weight of the neighboring level in the sum becomes smaller 
compared with the total weight of the far levels.

%%%%%%%%%%%%%%%%%%%%%%%%%%%%%%%%%%%%%%%%%%%%%%%%%%%%%%%%%%%%%%%%%%%%%%
%%%%%%%%%%%%%%%%%%%%%%%%%%%%%%%%%%%%%%%%%%%%%%%%%%%%%%%%%%%%%%%%%%%%%%
\section{Summary}
\label{S12}

We have developed a practical procedure for the 
analysis of a one dimensional double well system, 
which is both powerful and illuminating. 
The procedure assumes that we have a way to find 
the eign-energies $E_n$ of the device, 
and the mixing ratio $\Theta_n$ of each of them.  
Given the transmissions of the barriers  
we further characterize the device by 
the splitting ratio $\lambda$. With these ingredients 
in hand we can analyze any stirring process
and obtain explicit expressions for the geometric 
conductance~$G$. The calculation simplifies 
if the TLS modeling applies, because then the 
mixing ratio can be determined form 
the diagonalization of a $2\times 2$ matrix.

In particular we obtain explicit expressions 
for~$G$ due to either barrier translation
(generalizing a result that has been obtained in~\citeref{pmx}),    
or barrier modulation 
(generalizing a result that has been obtained in~\citeref{pms}),   
and verify that they agree 
with the naive self-consistent TLS calculation. 
We see that whenever the TLS modeling applies 
the proper calculation in the full Hilbert space  
gives the same result as the naive calculation 
in the TLS Hilbert space.

As a by product of the TLS analysis we find that 
the pumped ``charge" during an avoided crossing 
is not quantized (see Eq.(\ref{e44})), 
not only because of the topological splitting effect, 
but also due to a dynamical effect 
that arises if the barrier is modulated.

The practical importance of the TLS modeling 
in condense matter physics is obvious.
On the other hand the specific application to 
the study of quantum stirring deserves 
a few words regarding the measurement procedure 
and the experimental relevance. As explained  
in \citeref{pmx} it should be clear that the measurement  
of current in a closed circuit requires special 
techniques \citeref{orsay,expr1,expr2}. 
These techniques are typically used in order 
to probe persistent currents, 
which are zero order (conservative) effect, 
while in the present paper we were discussing driven 
currents, which are a first-order (geometric) effect. 
It is of course also possible to measure the dissipative 
conductance (as in~\citeref{orsay}).  
During the measurement the coupling to the system 
should be small. These are so called {\em weak measurement} 
conditions. More ambitious would be to measure 
the counting statistics, i.e. also the second moment 
of $Q$ as discussed in \citeref{cnb,cnz}  which is 
completely analogous to the discussion of noise measurements 
in open systems \citeref{levitov,nazarov}.
Finally it should be pointed out that the formalism 
above, and hence the results, might apply to 
experiments with superconducting circuits (see \citeref{JJ}).

%%%%%%%%%%%%%%%%%%%%%%%%%%%%%%%%%%%%%%%%%%%%%%%%%%%%%%%%%%%%%%
\appendix
\section{Conventions and notations}

Consider two segments that are connected 
at points that are labeled as $x_A$ and $x_B$.
In the absence of coupling each segment is 
regarded as a one dimensional box. 
The unperturbed eigenstates are labeled by~$i$ 
(or optionally by~$j$). In the TLS scheme ${i=1,2}$.
If the coupling is non-zero the exact eigenstates 
are labeled by~$n$ (or optionally by $m$).
Within the framework of the neighboring level 
approximation scheme we focus on two levels 
that we label as ${n=n_0}$ and ${m=m_0}$. 
If the TLS modeling applies then the states $n_0$ and $m_0$ 
are regarded as linear combinations of ${i=1,2}$. 

The unperturbed states ${i=1,2}$ are characterized 
by their parities $\pm^{\!\!1}$ and $\pm^{\!\!2}$ respectively. 
The relative sign $\pm\signC$ in Eq.(\ref{e7}) 
equals the product of $\pm^{\!\!1}$ and $\pm^{\!\!2}$.
Inverting the arbitrary gauge sign 
of either $\psi^{(1)}(x)$ or $\psi^{(2)}(x)$ 
would multiply the expression in Eq.(\ref{e333})  
by a global minus sign, while the relative 
sign $\pm\signC$ remains unchanged. 
The gauge invariant relative sign is due to the fact that 
the unperturbed states are either odd or even:
we have plus sign if both states have the
same parity and minus sign if they have opposite parity.

Each exact eigenfunction~$n$, as written as in Eq.(\ref{e37}),  
is characterized by what we call the parity $\pm\signA$ 
with respect to barrier~A. Positive parity means that 
the radial derivatives as defined in Eq.(\ref{e16}) 
have both the same sign. Optionally we can define $\pm^{\!\!b}$
as the parity with respect to barrier~B.
This parity $\pm\signA$ is not a symmetry related quantum number,  
but it is merely required in order to define the wavefunction of Eq.(\ref{e37}) 
in a unique way given the energy and the mixing ratio.
If the TLS modeling applies then for positive (negative) $\kappa$ 
the state $n_0$ of Eq.(\ref{e9}) has negative (positive) parity, 
while the $m_0$ state has positive (negative) parity.   
Within this framework the parity $\pm^{\!\!b}$ with respect 
to barrier~B is $\pm\signA$ multiplied by $\pm\signC$.

We have verified that the various $\pm$ signs through the 
paper are consistent, which is not always evident in a superficial look.

%%%%%%%%%%%%%%%%%%%%%%%%%%%%%%%%%%%%%%%%%%%%%%%%%%%%%%%%%%%%%%%%%%%%%%%%%%%
%%%%%%%%%%%%%%%%%%%%%%%%%%%%%%%%%%%%%%%%%%%%%%%%%%%%%%%%%%%%%%%%%%%%%%%%%%%

\acknowledgments

This research was supported by grants from 
the USA-Israel Binational Science Foundation (BSF), 
and from the Deutsch-Israelische Projektkooperation (DIP).

%%%%%%%%%%%%%%%%%%%%%%%%%%%%%%%%%%%%%%%%%%%%%%%%%%%%%%%%%%%%%%%%%%%%%%%%%%%
%%%%%%%%%%%%%%%%%%%%%%%%%%%%%%%%%%%%%%%%%%%%%%%%%%%%%%%%%%%%%%%%%%%%%%%%%%%

%%%%%%%%%%%%%%%%%%%%%%%%%%%%%%%%%%%%%%%%%%%%%%%%%%%%%%%%%%
%%%%%%%%%%%%%%%%%%%%%%%%%%%%%%%%%%%%%%%%%%%%%%%%%%%%%%%%%%

\clearpage

\begin{figure}[h]

\includegraphics[width=\hsize]{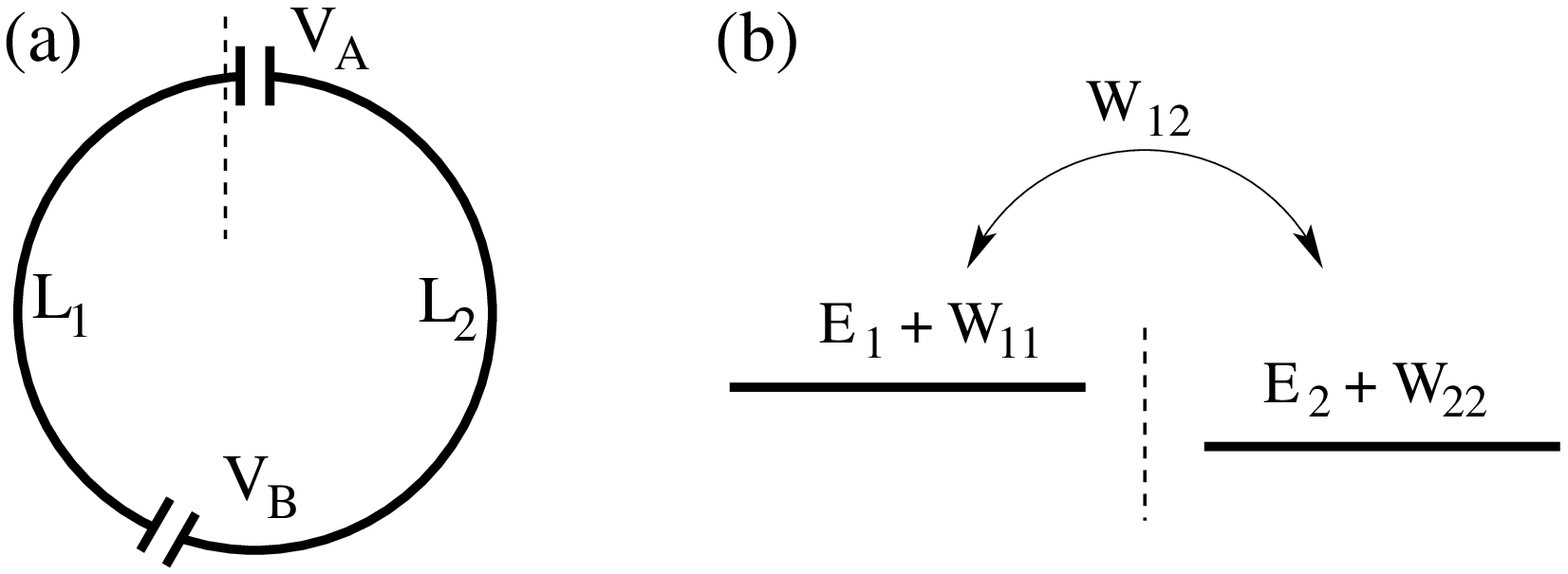}

\caption{
{\em Panel~(a)}: Illustration of a ring shaped device 
that is divided by the barriers~$V_A$ and~$V_B$ 
into two arms of length $L_1$ and $L_2$.
The current is measured through the dashed section 
near barrier~A.  In the quantum stirring scenario 
it is assumed that there is a gate control 
over the potential floor of each arm, 
or over the height or the location of barrier~B. 
{\em Panel~(b)}: 
Within the framework of the TLS modeling, the 
reduced Hilbert space contains two levels.
The perturbation $W_{ij}$ is due to having finite 
rather than infinite barriers, so it corresponds 
to the difference ${\mathcal{H}-\mathcal{H}(\infty)}$ 
and not to ${V = \mathcal{H}-\mathcal{H}(0)}$. 
See the text for further details.}
\end{figure}

\begin{figure}[h]

\includegraphics[width=0.6\hsize]{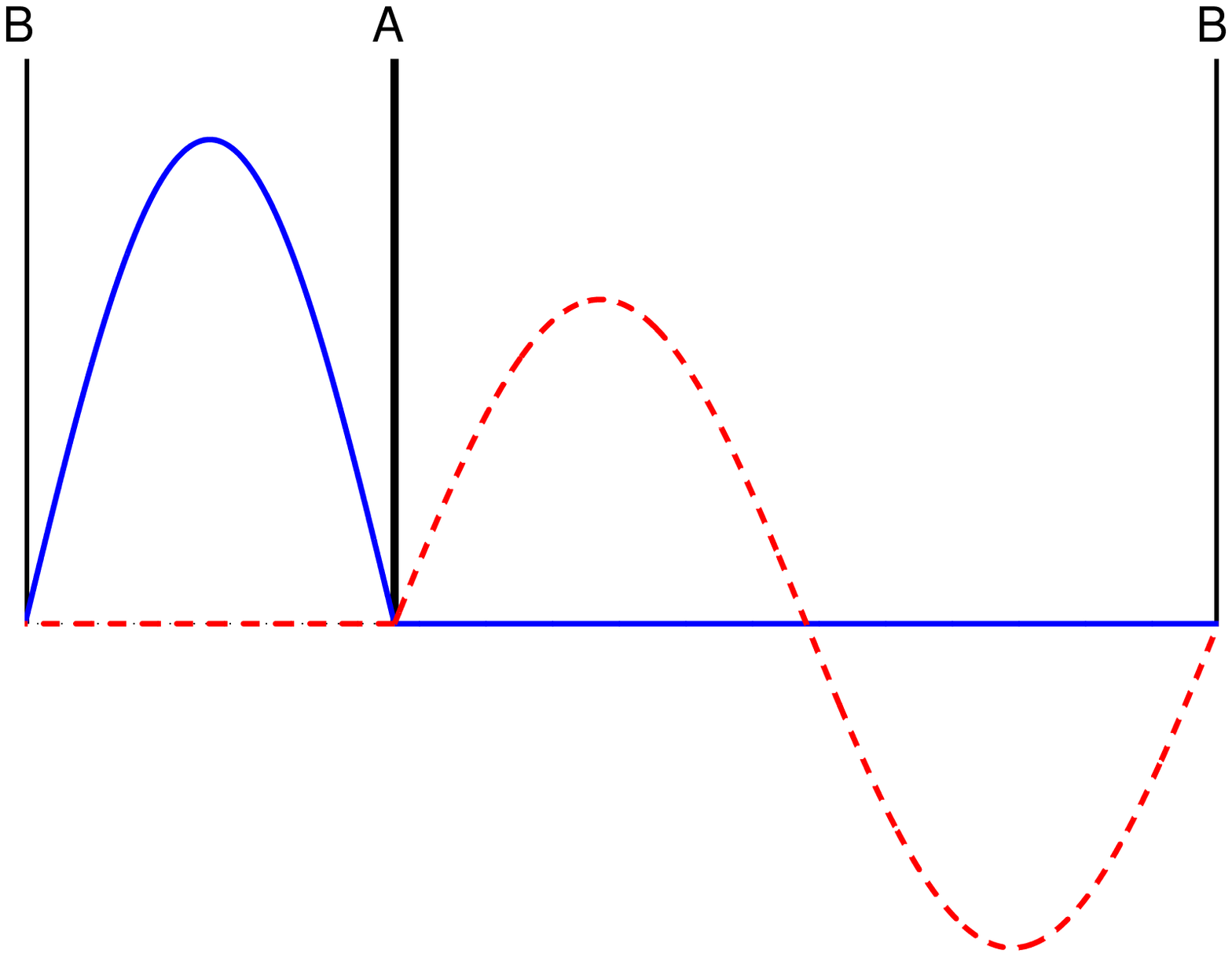}
\includegraphics[width=0.6\hsize]{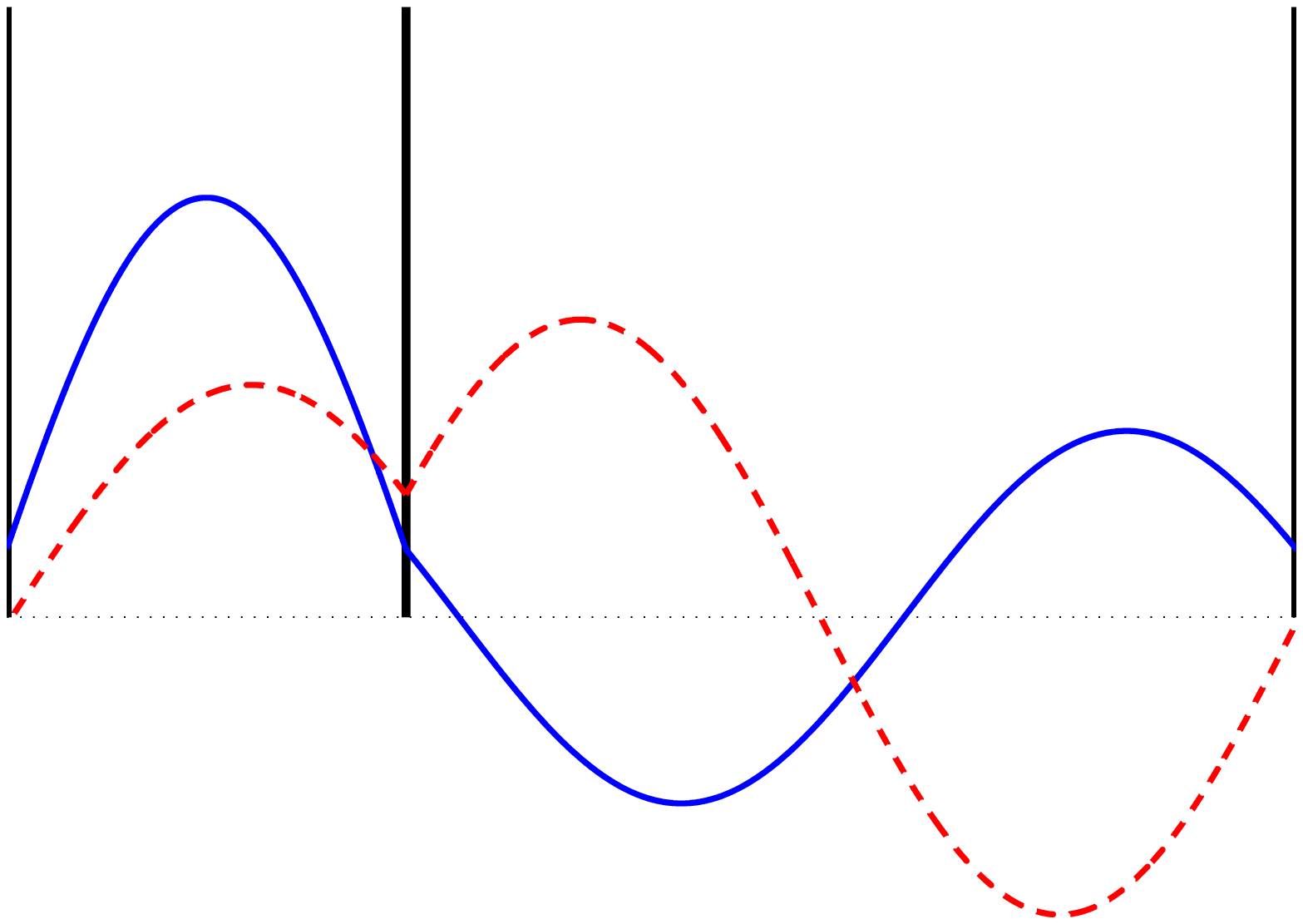}

\caption{
{\em Upper panel:} Two nearly degenerate eigenfunctions $\psi(x)$ of a 
particle in a ring with arms of length ${L_1=1}$ and ${L_2=2.23}$.
These are the two unperturbed states of Eq.(\ref{e4}).
{\em Lower panel:} The exact eigenfunctions 
assuming that the barriers are finite (${g_A\approx0.28}$ and ${g_B\approx0.06}$). 
These do not vanish at the barriers,  
and therefore cannot be written as a superposition of 
the unperturbed states. Still we explain in the text 
how a decent approximation for the former can be obtained 
using the neighboring levels approximation scheme.}
\end{figure}

\begin{figure}[h]

\includegraphics[width=0.7\hsize]{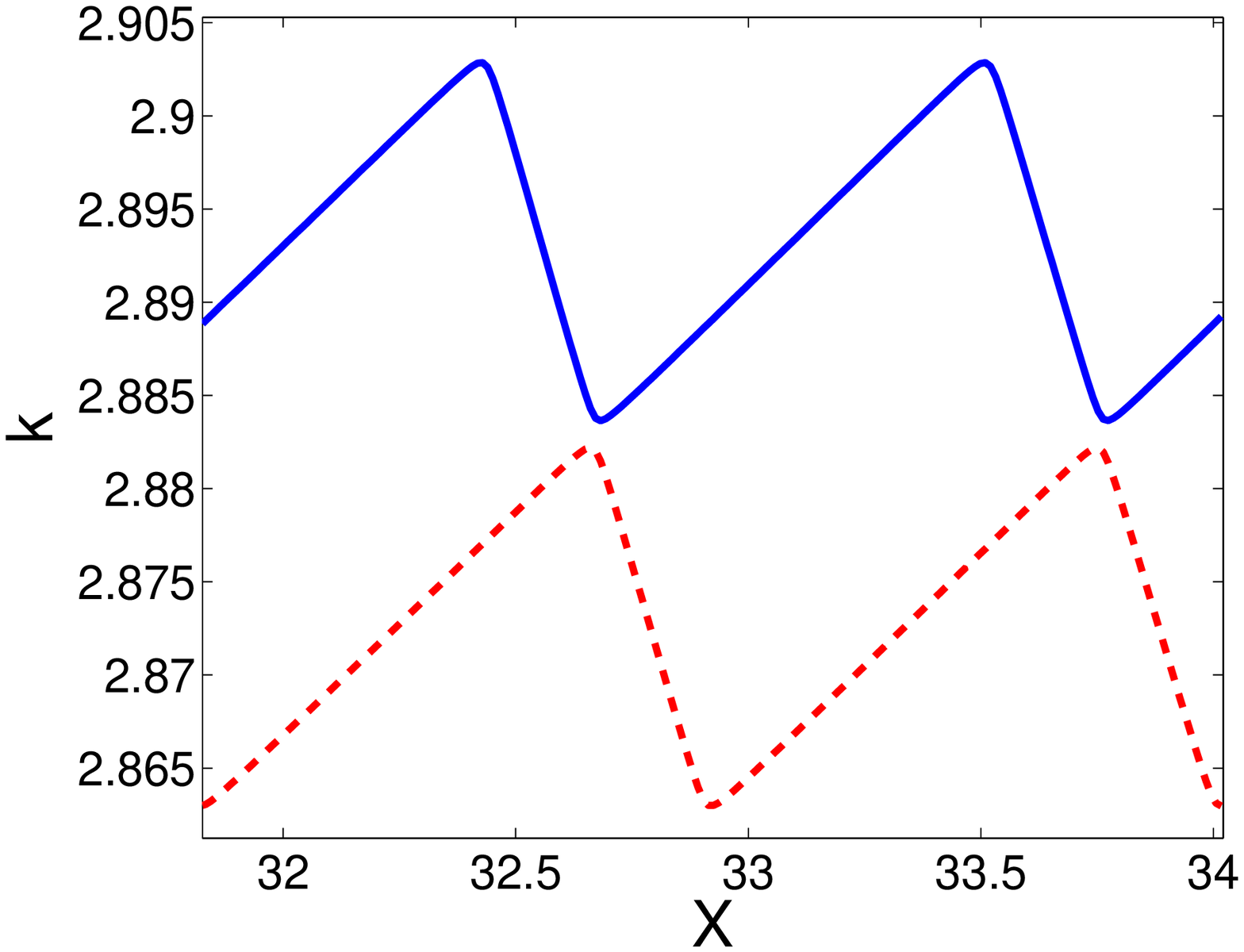}
\includegraphics[width=0.7\hsize]{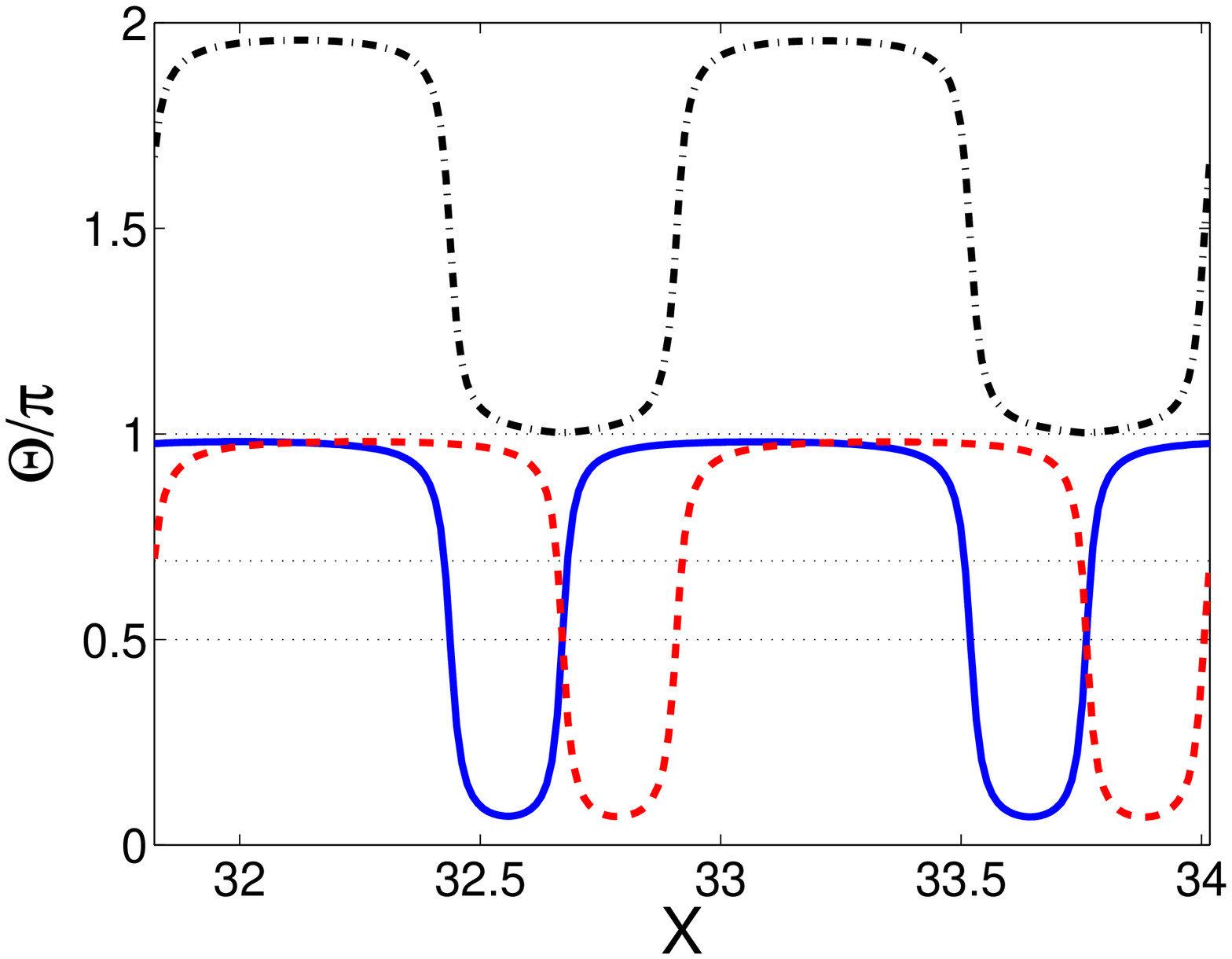}

\caption{
We consider a particle of mass ${\mass=1}$ in a ring 
of length $L=151.43$.  The position of barrier~B is $X$, 
so we have ${L_1=X}$ and ${L_2=L-X}$. 
We calculate numerically $k_n$ and $\Theta^{(n)}$ for two 
neighboring levels (solid and dashed lines). 
The sum $\Theta^{(m)}{+}\Theta^{(n)}$ is plotted as a dash-dotted line.
We have high barriers with ${g_A\sim10^{-2}}$ and ${g_B\sim10^{-5}}$.   
Accordingly we expect TLS modeling to be valid:
The dotted lines indicate the values ${\Theta=\pi/2}$ (expected crossing point) 
and ${\Theta=\pi}$ (expected sum). For sake of comparison 
there is a third dotted line that indicates 
the value of $\Theta$ that corresponds to equal amplitudes ${C_1=C_2}$.
}
\end{figure}

\begin{figure}[h]

\includegraphics[width=0.49\hsize]{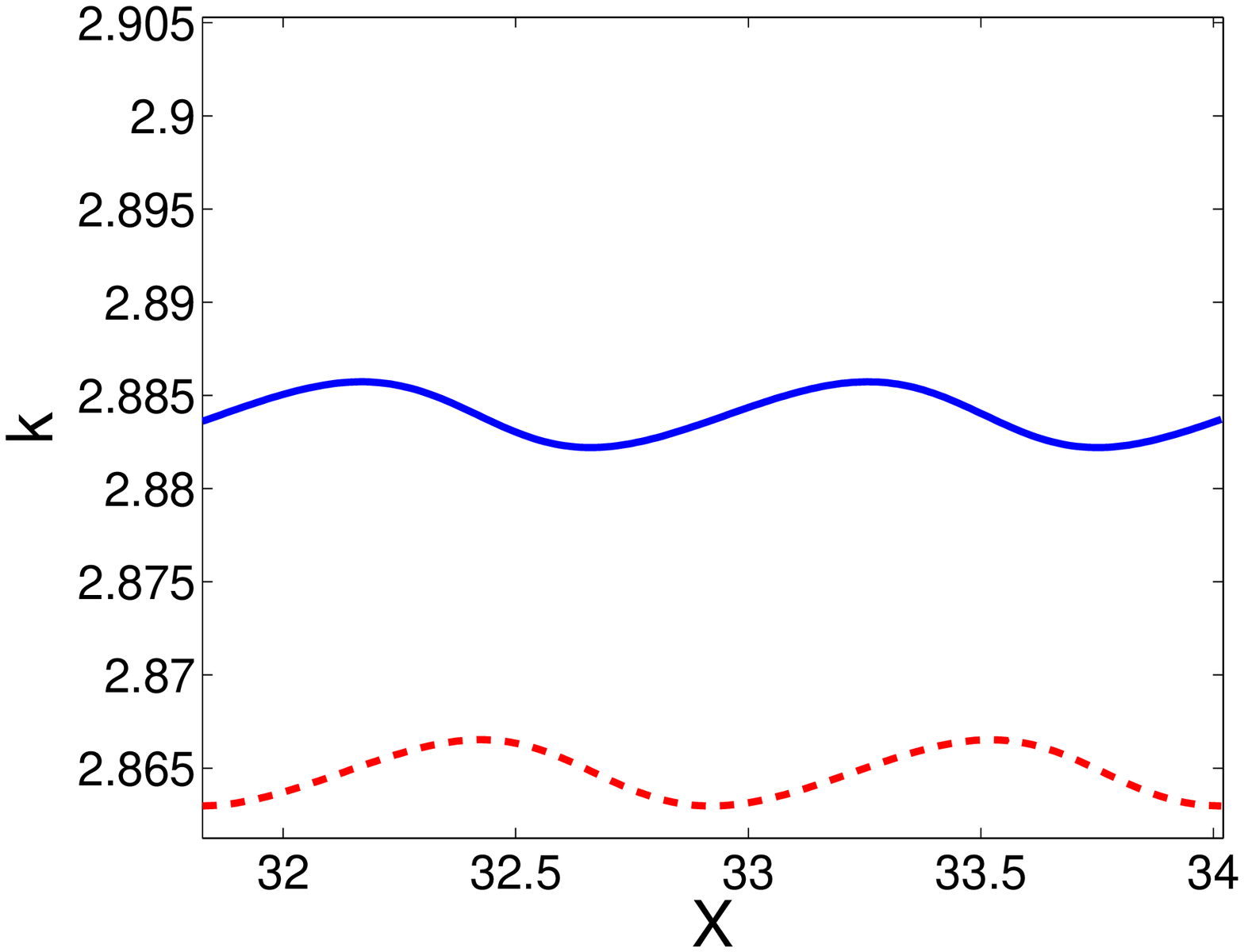}
\includegraphics[width=0.49\hsize]{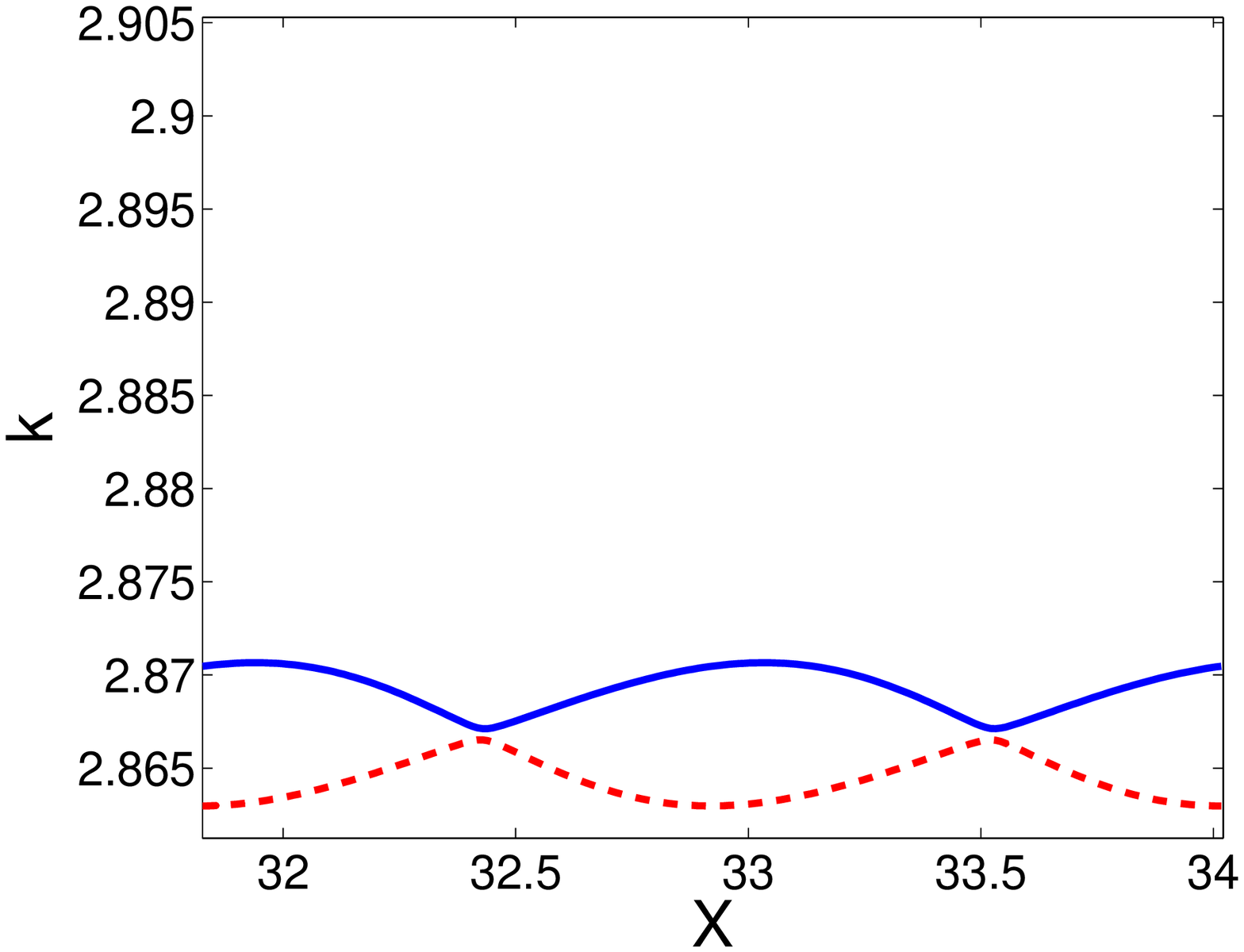}
\\
\includegraphics[width=0.49\hsize]{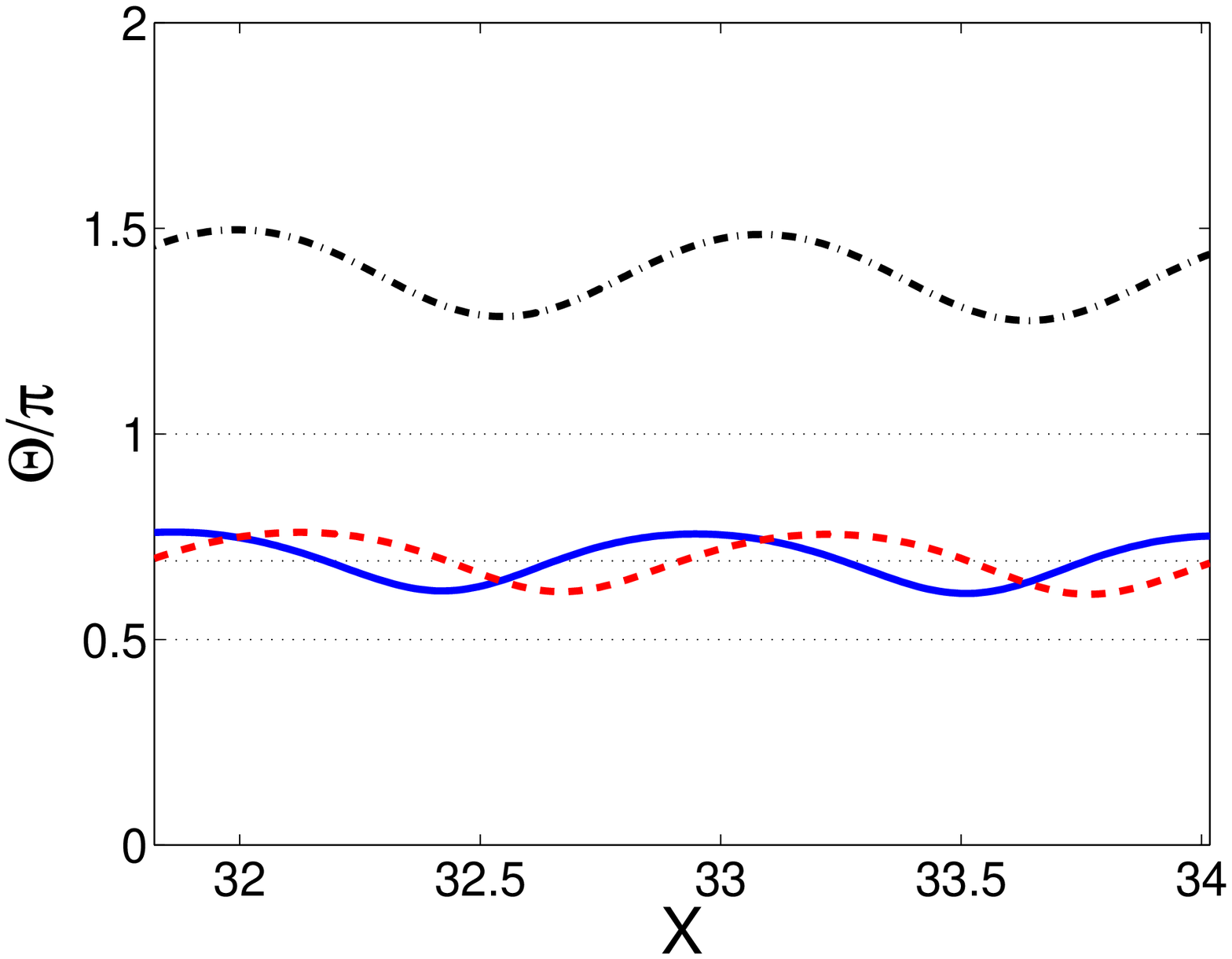}
\includegraphics[width=0.49\hsize]{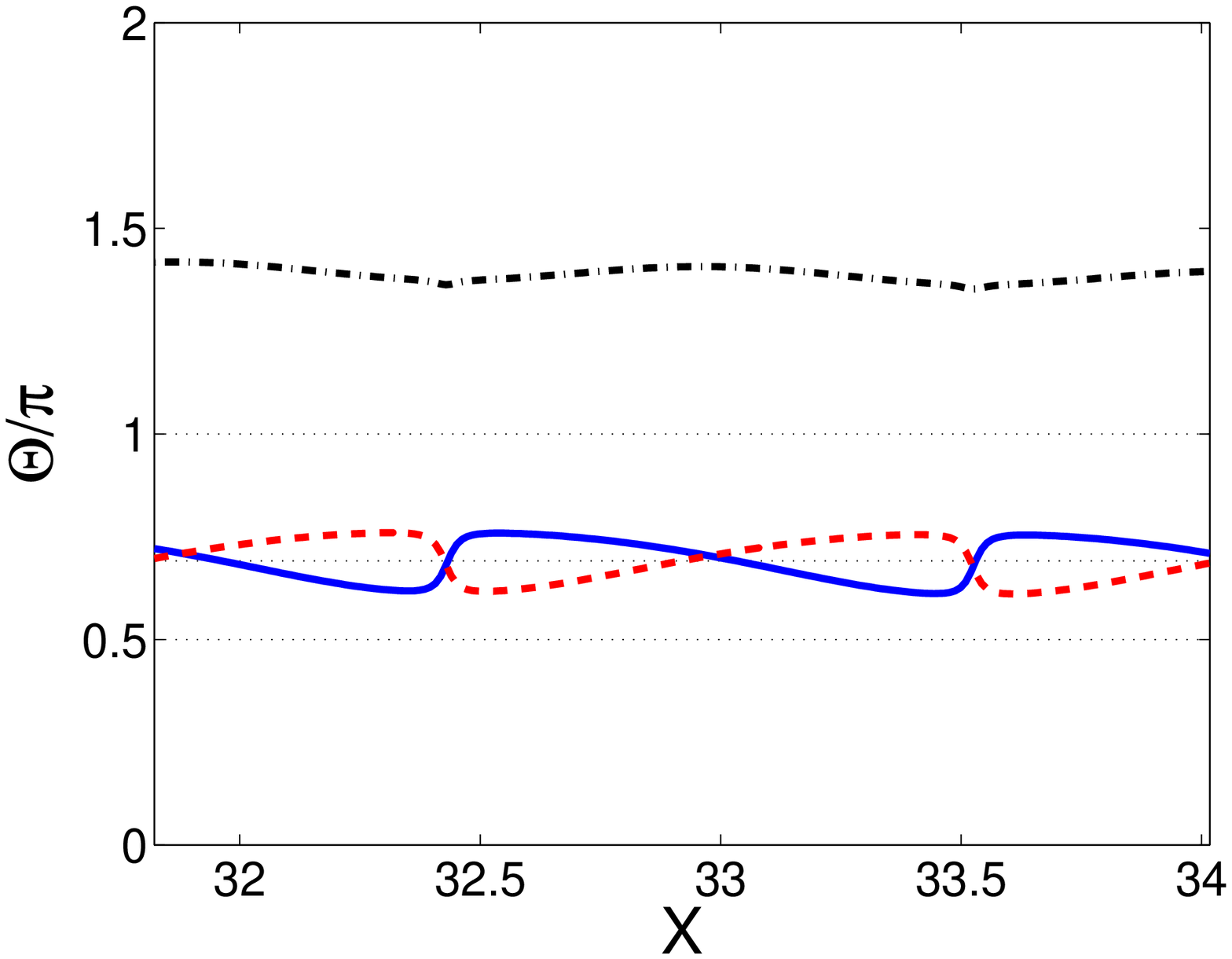}

\caption{
The same as the previous figure, but here 
the TLS modeling does not apply. 
In the left panels 
one barrier is high (${g_A\sim10^{-2}}$) 
and one barrier is low (${g_B\sim0.9}$), 
while in the right panels 
both barrier are low (${g_A\sim g_B\sim0.9}$).
}
\end{figure}

\clearpage

%%%%%%%%%%%%%%%%%%%%%%%%%%%%%%%%%%%%%%%%%%%%%%%%%%%%%%%%%%%%%%%%
%%%%%%%%%%%%%%%%%%%%%%%%%%%%%%%%%%%%%%%%%%%%%%%%%%%%%%%%%%%%%%%%

\newpage
\begin{thebibliography}{99}


%%Quantum stirring

\bibitem{pmx}
The most recent publication in this line of study is:    
I. Sela and D. Cohen, Phys. Rev. B {\bf 77}, 245440 (2008).
For older references see there. 


%%Quantum pumping

\bibitem{Thouless}
D. J. Thouless, Phys. Rev. B {\bf 27}, 6083 (1983).

\bibitem{bpt} 
M. Buttiker, H. Thomas and A. Pretre, Z. Phys. B {\bf 94}, 133 (1994).

\bibitem{BPT2}
P. W. Brouwer, Phys. Rev. B {\bf 58}, 10135 (1998). 

\bibitem{Avron}
J. E. Avron, A. Elgart, G. M. Graf and L. Sadun,  Phys. Rev. B {\bf 62}, 10618 (2000).

\bibitem{pMB}
M.~Moskalets and M.~B{\"u}ttiker, 
Phys. Rev. B {\bf 68}, 161311 (2003).

\bibitem{pms} 
I. Sela and D. Cohen, J. Phys. A {\bf 39}, 3575 (2006). 



%%Kubo

\bibitem{pmc}
D. Cohen, Phys. Rev. B {\bf 68}, 155303 (2003).

\bibitem{berry1}
M.V. Berry, Proc. R. Soc. Lond. A {\bf 392}, 45 (1984). 

\bibitem{avron2}
J. E. Avron, A. Raveh and B. Zur, Rev. Mod. Phys. {\bf 60}, 873 (1988). 

\bibitem{berry2}
M.V. Berry and J.M. Robbins, Proc. R. Soc. Lond. A {\bf 442}, 659 (1993).




% measurement

\bibitem{orsay} 
Measurements of currents in arrays 
of closed rings are described by:  \
B. Reulet M. Ramin, H. Bouchiat and D. Mailly, 
Phys. Rev. Lett. {\bf 75}, 124 (1995). 

\bibitem{expr1} 
Measurements of currents in individual closed rings 
using SQUID is described in: \
N.C. Koshnick, H. Bluhm, M.E. Huber, K.A. Moler, Science 318, 1440 (2007).

\bibitem{expr2} 
A new micromechanical cantilevers technique for measuring currents 
in closed rings is described in:  
A.C. Bleszynski-Jayich, W.E. Shanks, R. Ilic, J.G.E. Harris, arXiv:0710.5259,
Journal of Vacuum Science \& Technology B {\bf 26}, 1412 (2008).


%measure

\bibitem{cnb}
M. Chuchem and D. Cohen, J. Phys. A {\bf 41}, 075302  (2008).

\bibitem{cnz}
M. Chuchem and D. Cohen, Phys. Rev. A {\bf 77}, 012109 (2008).

\bibitem{levitov}
L.S. Levitov and G.B. Lesovik, JETP Letters {\bf 58}, 230 (1993).

\bibitem{nazarov}
Y.V. Nazarov and M. Kindermann, European Physical Journal B {\bf 35}, 413 (2003).

\bibitem{JJ}
M. Mottonen, J. P. Pekola, J. J. Vartiainen,
V. Brosco and F. W. J. Hekking, Phys. Rev. B {\bf 73}, 214523 (2006).



\end{thebibliography}
\end{document}